\documentclass[twocolumn,aps,showpacs,floatfix,prc]{revtex4}
\usepackage{graphicx}
\begin{document}
\title{ Binding, bonding and  charge symmetry breaking  in  $\Lambda$-hypernuclei 
 }

\author{ Chhanda Samanta, Thomas A. Schmitt }

\affiliation{Department  of Physics $\&$ Astronomy, Virginia Military Institute, Lexington, VA 24450, USA }


\date{\today }

\begin{abstract}

Recent experiments have presented more accurate data on the $\Lambda\Lambda$-binding energies of a few $\Lambda\Lambda$-hypernuclei. This is important as the $\Lambda\Lambda$-bond energies ($\Delta$ΔB$_{\Lambda\Lambda}$) of double-$\Lambda$ hypernuclei provide a measure of the in-medium strength of the $\Lambda\Lambda$-interaction. A mass formula, optimized with the newly available $\Lambda\Lambda$ binding energy data, is used to estimate the binding energy and bond energy over a wide range of hypernuclei. The $\Delta$ΔB$_{\Lambda\Lambda}$ values calculated with this mass formula are in good agreement with the experimental data and the  predictions of the quark mean-field (QMF) and relativistic mean-field (RMF) models, except at low mass region where large uncertainties exist in the current experimental data. The $\Lambda\Lambda$-bond energies in $\Lambda\Lambda$-hypernuclei are found to diminish with neutron numbers, approaching zero near the neutron-drip line. In this formalism, the calculated binding energy difference in mirror nuclei arises from the Coulomb contributions and can be utilized to extract the Coulomb-corrected charge symmetry breaking effect in mirror $\Lambda$-hypernuclei. Our calculations show the regions where more experimental data are needed for light and neutron-rich $\Lambda$ and $\Lambda\Lambda$-hypernuclei.


\end{abstract}

\maketitle

\noindent
\section{Introduction}
\label{section1}

The discovery of a hypernucleus in the early nineteen-fifties~\cite{DP53} ushered in a new era in nuclear physics. Recently much attention has been focused on the neutron-rich hypernuclei with Lambda ($\Lambda$)-hyperons~\cite{TA14,GO16R}. While the nucleon-nucleon interaction is well studied, the knowledge of $\Lambda$-neutron,  $\Lambda$-proton and  $\Lambda$-$\Lambda$ interaction is still evolving.  

One interesting question is how two $\Lambda$-hyperons would interact with each other while they are bound inside a nucleus.  It has  been found that in some hypernuclei the $\Lambda\Lambda$-separation energy ($B_{\Lambda\Lambda}$) from a double-$\Lambda$ hypernucleus exceeds twice the value of a single-$\Lambda$  separation energy ($B_{\Lambda}$) of a single-$\Lambda$ hypernucleus, while both the $\Lambda$ and $\Lambda\Lambda$-hypernuclei have the same core nucleus (i.e., same number of neutrons and protons). This excess, $\Delta B_{\Lambda\Lambda}$,  is known as the $\Lambda\Lambda$-bond energy (or, interaction energy)~\cite{DM97}.  Experimentally $\Delta B_{\Lambda\Lambda}$ is found to be larger for light nuclei than the heavier nuclei, although its variation with the neutron numbers for nuclei with different proton numbers has not been studied so far. Therefore, we do not know to what extent the $\Lambda\Lambda$-bonding will persist in neutron-rich hypernuclei.

 Another existing puzzle is the large charge symmetry breaking (CSB) effect observed in light mirror hypernuclei~\cite{BO85,GA06,GAL15,GAL16,AC17,BOT17}. In normal nuclear systems, without any hyperons,  the charge symmetry breaking (CSB) in strong interaction occurs because of the mass difference of the up (u) and down (d) quarks of neutrons (udd) and protons (uud)~\cite{GA06}. In the mirror non-strange A=3 nuclei, $^3$He(p+p+n)and $^3$H(p+n+n), a binding-energy difference of 0.7638 $\pm$0.0003 MeV was observed of which approximately 0.081 MeV was attributed to the CSB  effect after making the 0.683 MeV Coulomb correction~\cite{GO16R,BO85,GAL15}. 
 
 The $\Lambda$-hyperon has one up, one down and one strange quark in it, and it has no isospin or charge. 
 The charge symmetry of strong interaction dictates that in mirror hypernuclei the $\Lambda$p and $\Lambda$n interactions and their contributions in the $\Lambda$-binding energy should be identical. Comparing  the ground state binding energies of  A=4 mirror hypernuclei $^4_{\Lambda}$He (p+p+n+$\Lambda$) and $^4_{\Lambda}$H (p+n+n+$\Lambda$), a large $\Delta$B$_\Lambda$($^4_{\Lambda}$He~-~$^4_{\Lambda}$H) = B$_\Lambda(^4_{\Lambda}$He)~-~B$_{\Lambda}(^4_{\Lambda}$H) = (2.39$\pm$0.03)~-~(2.04$\pm$0.04) = +0.35 $\pm$0.05 MeV is found~\cite{GO16R}. Although it is known that addition of a $\Lambda$-hyperon would contract a nucleus, the exact estimation of the Coulomb correction in hypernuclei is difficult~\cite{GAL15, AC17,GAL16,BOT17}. The Coulomb energy was predicted to be $\sim$0.02-0.08 MeV that led to $\Lambda$-hypernuclear  CSB effect to be $\Delta$B$_\Lambda$($^4_{\Lambda}$He~-~$^4_{\Lambda}$H) + $\Delta$B$_c\sim$0.4 MeV for the above A=4 isodoublet hypernuclear system. Thus the CSB effect in the A=4 mirror pair $^4_{\Lambda}$He (=$^3$He + $\Lambda$), and  $^4_{\Lambda}$H (=$^3$H + $\Lambda$) is approximately five times larger than that for the A = 3 non-strange mirror pair $^3$He and $^3$H.

For determination of both the $\Lambda\Lambda$-bond energy in $\Lambda\Lambda$-hypernuclei and the charge symmetry breaking effect in mirror $\Lambda$-hypernuclei, it is necessary to have very accurate experimental data on the $\Lambda$ and $\Lambda\Lambda$-binding energies. Hence, many old emulsion data are now being revisited using high energy accelerators coupled with powerful spectroscopic tool. For example, the spectroscopy of $^{10}_{\Lambda}$Be \cite{GO16} and $^{7}_{\Lambda}$He \cite{GO16R} was carried out at JLab, using the (e,e'k$^+$) reaction. Both experiments provided significantly  more accurate values for $B_{\Lambda}$ than that of the previous emulsion studies.

In this work the above problems have been addressed using an optimized mass formula ~\cite{SAM18,SAM10,SAM06} that reproduces the currently available experimental data of the $B_{\Lambda}$ and $B_{\Lambda\Lambda}$~\cite{BOT17,BA90,FENA15,NA15,NA14,NA10,DA05} which is then used to compute the $\Delta B_{\Lambda}$ and $\Delta B_{\Lambda\Lambda}$ in hypernuclei over a wide mass range.  The ${\Lambda\Lambda}$-bond energy is found to diminish rapidly with the increasing mass number, in agreement with the experimental data. The value of  $\Delta B_{\Lambda\Lambda}$ is significantly larger for  nuclei with lower proton (Z) numbers than the higher ones. Interestingly, the calculated $\Delta B_{\Lambda\Lambda}$ is found to diminish rapidly with increasing neutron number, even for low Z nuclei. But, as mentioned before, no experimental data exists for such neutron-rich nuclei.

 In the formalism considered in this work, the CSB effect cannot be computed. Instead, the difference in $\Lambda$-binding energies in mirror nuclei arises from the difference in Coulomb energies, a quantity that is otherwise hard to predict for a bound hypernucleus. Hence, this mass formula  can be effectively utilized to extract the Coulomb-corrected CSB effect from  the experimental data of mirror nuclei.

\noindent
\section{Formalism}
\label{section2}

The $ B_{\Lambda}$, $ B_{\Lambda\Lambda}$  and $\Delta B_{\Lambda\Lambda}$  of an element with Z number of protons and N number of neutrons are given as, \\

\begin{equation} \label{eq1}
 B_{\Lambda}(^A_{\Lambda}Z ) = M(^{A-1} Z)  + M({\Lambda}) - M( ^A_{\Lambda}Z)
\end{equation}

\begin{equation} \label{eq2}
B_{\Lambda\Lambda}(^A_{\Lambda\Lambda}Z ) = M(^{A-2}Z)  + 2M({\Lambda})  - M(^A_{\Lambda\Lambda}Z)  	
\end{equation}
	 
\begin{equation}  \label{eq3}
{\Delta}B_{\Lambda\Lambda}(^A_{\Lambda\Lambda}Z ) = B_{\Lambda\Lambda} (^A_{\Lambda\Lambda}Z) - 2B_{\Lambda}(^{A-1}_{\Lambda}Z)	              
\end{equation}

\vspace{0.5cm}
\noindent where $M(\Lambda)$ is the mass of the $\Lambda$  hyperon, $M(^{A-1} Z)$ is the mass of the core nucleus of the single-$\Lambda$ nucleus of mass $ M( ^A_{\Lambda}Z)$, and $M(^{A-2}Z)$  is the mass of the core nucleus of the double-$\Lambda$  nucleus of mass  $M(^A_{\Lambda\Lambda}Z) $. Here, the mass number (A) of a hypernucleus is the total number of baryons (neutrons, protons and hyperons) in the nucleus. 

The bond energies ($\Delta B_{\Lambda\Lambda}$) are the measures of the energies released when the $\Lambda\Lambda$ bond is broken. These energies assist in understanding the nature of the in-medium strength of the~$\Lambda$-$\Lambda$ interaction. Earlier,  the ${\Delta}B_{\Lambda\Lambda}$-values of a few nuclei were calculated in a microscopic framework using the Quark Mean Field (QMF) and Relativistic Mean Field (RMF) models~\cite{HU14}. In the QMF model the  $\Lambda$ and $\Lambda\Lambda$-hypernuclei were studied with broken SU(3) symmetry for the quark confinement potential. Such symmetry breaking improved the description of the binding energies of $\Lambda$ and $\Lambda\Lambda$ hypernuclei. The microscopic calculations being complicated and time consuming, it is difficult to carry out calculations for many nuclei at the same time. Whereas a mass formula can give a quick estimation of the binding energies of a large number of nuclei in a very short time.

The pair of hypernuclei with the same total number of baryons  but with the neutron  and proton  numbers interchanged, are called mirror hypernuclei.  The $\Delta B_{\Lambda}$  is calculated as,

\begin{equation}  \label{eq4}
\Delta B_{\Lambda}= B_{\Lambda}(^{ N+Z+\Lambda}_{\Lambda} Z ) - B_{\Lambda}(^{Z + N + \Lambda}_{\Lambda} Z^{\prime})             
\end{equation}
\vspace{0.1cm}

\noindent where  $Z^{\prime}$ is the number of protons in the second nucleus and is equal to the number of neutrons of the first nucleus with proton number Z.  Both nuclei have the same total number of baryons (Z+N+$\Lambda$). Owing to the difference in the proton numbers of the two hypernuclei, the Coulomb contributions in the binding energies are different. The CSB effect for hyperon-nucleon interaction is extracted from the $\Delta B_{\Lambda}$ after careful consideration of the difference in Coulomb effects that arises from the difference in proton numbers.

In this work, the $B_\Lambda$ and $B_{\Lambda\Lambda}$ have been calculated using an optimized generalized mass formula ~\cite{SAM18,SAM10,SAM06} that is applicable to both non-strange nuclei and strange hypernuclei of mass number A= N + Z$_c$ + n$_Y$, containing N number of neutrons, Z$_c$ number of protons and  n$_Y$  number of hyperons (Y).  This formula is applicable to Lambda ($\Lambda$) as well as cascade ($\Xi$) -hypernuclei due to explicit consideration of the charge number q$_Y$ (with proper sign), mass m$_Y$ (in MeV) and strangeness S of each kind of hyperon. In this formalism,  the hypernucleus is considered as a core of normal nucleus plus the hyperon(s) and the binding energy is defined as~\cite{SAM18},

\begin{eqnarray}  
 B (A, Z) = 15.777A - 18.34A^{2/3} - 0.71Z(Z-1)/A^{1/3} \nonumber\\
- 23.21(N-Z_c )^2/[(1+e^{-A/17} )A] \nonumber \\
+ n_Y [0.0335 m_Y - 27.8 - 48.7|S|/A^{2/3} ] + \delta.  
\label{eq5}
\end{eqnarray}

\noindent 
The total charge of the hypernucleus $Z$  is  given by,     
\begin{equation}
Z = Z_c + n_Y q_Y.
\label{eq6}
\end{equation}
\noindent
 The strangeness  of the $\Lambda$ hyperon is  S = 1 and the mass is taken as, m$_Y$ =1115.683 MeV.
For non-strange (S=0) normal nuclei $n_Y$ = 0; hence $Z=Z_c$. For $\Lambda$-hypernuclei the charge $q_Y$ = $q_{\Lambda}$ = 0 that leads to  $Z = Z_c$ as well. It may be mentioned here that due to recent availability of  more accurate $\Lambda\Lambda$-separation energy values of a few double-$\Lambda$ hyperuclei, the old parameter 26.7 used earlier in the above binding energy formula~\cite{SAM06} has been changed to 27.8. This minor modification better reproduces both the $\Lambda$- and $\Lambda\Lambda$-separation energies, especially near the low mass regions. 

The pairing term $\delta$ of Eqn.5 is given by,

\begin{eqnarray}
 \delta=&&12A^{-1/2}(1-e^{-A/30})~for~even~ N-even~Z_c~nuclei,\nonumber\\
&&=-12A^{-1/2}(1-e^{-A/30})~for~ odd~N-odd~Z_c~nuclei, \nonumber\\
&&=0~ when~ N+Z_c~ is~ odd.\nonumber\\
\label{eq7}
\end{eqnarray} 
\noindent
The choice of $\delta$ value depends only on the number of neutrons and protons ($Zc$) in both normal and hypernuclei.

The hyperon separation energies ($B_Y$) is expressed as,\\

\begin{equation}  \label{eq9}
B_Y =B(A,Z)_{hyper} - B(A - n_Y, Z_c)_{core}				
\end{equation}
\vspace{0.3cm}

\noindent  where $B(A,Z)_{hyper}$ is the binding energy of the hypernucleus with mass number A = N + Z + $n_Y$, and $B(A - n_Y, Z_c)_{core}$ is the binding energy of its core nucleus without the hyperon(s). For a single-$\Lambda$ hypernucleus the number of hyperons is one ($n_Y$ = $n_{\Lambda}$= 1), and for a double-$\Lambda$ hypernucleus the number of hyperons is two ($n_Y$ = $n_{\Lambda}$ = 2).

\noindent

\noindent
\section{Results}
\label{section3}

Equation 8 has been used to calculate the $\Lambda$ and $\Lambda\Lambda$-separation energies. The  available experimental data on $B_{\Lambda}$ of different hypernuclei  and the results of this work (SAM)  are presented in the Table 1. The references of the experimental data are given inside the table in the last column. The calculated binding energy values are in good agreement with the experimental data~\cite{GO16R,BOT17,SAM06,BA90,HU14,HAS96}.

\begin{table*}
\caption{Comparison of the experimental values for the $\Lambda$-separation energies ($B_\Lambda$) in MeV,  of different $\Lambda$-hypernuclei with the results of the generalized mass formula (SAM). Statistical and systematic errors are written within brackets.}
\label{sphericcase}
\begin{tabular*}{\textwidth}{@{\extracolsep{\fill}}lrrrrl@{}}
\hline
\hline
Nucleus & \multicolumn{1}{c}{ $B_{\Lambda}$} & \multicolumn{1}{c}{$B_{\Lambda}$} & \multicolumn{1}{c}{$Error$} & \multicolumn{1}{c}{$Ref.$}\\
$^A_{\Lambda}$Z&(SAM) & Exp & (stat, sys)\\
\hline
\hline
$^{	208}_\Lambda$Pb	&	24.18	&	26.30	&	0.80	&	\cite{HU14}	\\
$^{	208}_\Lambda$Tl	&	24.24	&	 -	&	-	&	-	\\
$^{	139}_\Lambda$La	&	22.93	&	23.80	&	1.00	&	\cite{HAS96}	\\
$^{	139}_\Lambda$Ba	&	23.00   &	- 	    &	-	&	-	\\
$^{	91}_\Lambda$Zr	&	21.40	&	-    	&	-	&	-	\\
$^{	91}_\Lambda$Y	&	21.49	&	 -	    &	-	&	-	\\
$^{	89}_\Lambda$Y	&	21.35	&	23.10	&	0.50	&	\cite{HU14}	\\
$^{	89}_\Lambda$Sr	&	21.42	&	 -	    &	-	&	-	\\
$^{	56}_\Lambda$Fe	&	19.60	&	21.00	&	1.50	&	\cite{SAM06}	\\
$^{	56}_\Lambda$Mn	&	19.65	&	- 	&	-	&	-	\\
$^{	51}_\Lambda$V	&	19.27	&	19.90	&	1.00	&	\cite{SAM06}	\\
$^{	51}_\Lambda$Ti	&	19.35	&	- 	&	-	&	-	\\
$^{	41}_\Lambda$Ca	&	18.34	&	19.24	&	0	&	\cite{SAM06}	\\
$^{	41}_\Lambda$K	&	18.32	&	- 	&	-	&	-	\\
$^{	40}_\Lambda$Ca	&	18.28	&	18.70	&	1.10	&	\cite{HU14}	\\
$^{	40}_\Lambda$K	&	18.21	&	 -	&	-	&	-	\\
$^{	33}_\Lambda$S	&	17.34	&	17.96	&	0	&	\cite{SAM06}	\\
$^{	33}_\Lambda$P	&	17.32	&	- 	&	-	&	-	\\
$^{	32}_\Lambda$S	&	17.24	&	17.50	&	0.50	&	\cite{BA90}	\\
$^{	32}_\Lambda$P	&	17.16	&	- 	&	-	&	-	\\
$^{	28}_\Lambda$Si	&	16.56	&	16.00	&	0.29	&	\cite{SAM06}	\\
$^{	28}_\Lambda$Al	&	16.48	&	- 	&	-	&	-	\\
$^{	17}_\Lambda$O	&	13.52	&	13.59	&	0	&	\cite{SAM06}	\\
$^{	17}_\Lambda$N	&	13.58	&	 -	&	-	&	-	\\
$^{	16}_\Lambda$O	&	13.17	&	13.02	&	(0.5,0.08)	&	\cite{BOT17}	\\
$^{	16}_\Lambda$N	&	13.09	&	13.76	&	0.16	&	\cite{BOT17}	\\
$^{	15}_\Lambda$N	&	12.59	&	13.80	&	(0.7,1.0)	&	\cite{BOT17}	\\
$^{	15}_\Lambda$C	&	12.79	&	- 	    &	-	        &	-	\\
$^{	14}_\Lambda$N	&	12.21	&	12.17	&	0       	&	\cite{SAM06} 	\\
$^{	14}_\Lambda$C	&	12.12	&	12.17	&	(0.33,0.04)	&	\cite{BOT17}	\\
$^{	13}_\Lambda$C	&	11.55	&	11.98	&	(0.05,0.08)	&	\cite{BOT17}	\\
$^{	13}_\Lambda$B	&	11.72	&	 -	    &	-	        &	-	\\
$^{	12}_\Lambda$C	&	11.01	&	10.76	&	(0.19,0.04)	&	\cite{BOT17}	\\
$^{	12}_\Lambda$B	&	10.92	&	11.37   &	(0.06,0.04)     &	\cite{BOT17}]\\
$^{	11}_\Lambda$B	&	10.09	&	10.28	&	(0.2,0.4)	&	\cite{BOT17}	\\
$^{	11}_\Lambda$Be	&	10.51	&	-   	&	- 	            &	-	\\
$^{	10}_\Lambda$B	&	9.45	&	8.70	&	(0.1,0.08)	&	\cite{BOT17}	\\
$^{	10}_\Lambda$Be	&	9.36	&	9.11	&	(0.22,0.04) &	\cite{GO16}	\\
$^{	9}_\Lambda$B	&	8.99	&	8.29	&	(0.18,0.04)	&	\cite{BOT17}	\\
$^{	9}_\Lambda$Be	&	8.31	&	6.59	&	(0.07,0.08)	&	\cite{BOT17}	\\
$^{	9}_\Lambda$Li	&	8.79	&	8.53	&	0.15	    &	\cite{BOT17}	\\
$^{	8}_\Lambda$Be	&	7.33	&	6.84	&	(0.03,0.04)	&	\cite{BOT17}	\\
$^{	8}_\Lambda$Li	&	7.23	&	6.80	&	(0.03,0.04)	&	\cite{BOT17}	\\
$^{	8}_\Lambda$He	&	8.84	&	7.16	&	(0.70,0.04)	&	\cite{BOT17}	\\
$^{	7}_\Lambda$Be	&	6.89	&	5.16	&	(0.08,0.04)	&	\cite{BOT17}	\\
$^{	7}_\Lambda$Li	&	5.55	&	5.82	&	(0.08,0.08)	&	\cite{BOT17}	\\
$^{	7}_\Lambda$He	&	6.69	&	5.55	&	(0.10,0.11)	&	\cite{BOT17}	\\
$^{	6}_\Lambda$Li	&	4.21	&	4.50	&	0       	&	\cite{BA90} 	\\
$^{	6}_\Lambda$He	&	4.11	&	4.18	&	(0.10,0.04)	&	\cite{BOT17}	\\
$^{	5}_\Lambda$He	&	1.43	&	3.12	&	(0.02,0.04)	&	\cite{BOT17}	\\
\hline
\end{tabular*}
\end{table*}


The  $\Lambda\Lambda$-separation energies (in MeV) of different $\Lambda\Lambda$-hypernuclei  calculated in this work and the experimental data for $\Lambda\Lambda$-hypernuclei~\cite{GAL11} are presented in Table 2. The mass formula results are found to be in good agreement with the experimental data for $\Lambda\Lambda$-separation energies. The shell model predictions ({$B^{SM}_{\Lambda\Lambda}$}) of Ref.~\cite{GAL11} are also shown in the table for comparison. 

In Fig.1, the results of the generalized mass formula (SAM) are compared with the predictions of the liquid drop model (LD)~\cite{BOPO07,BUY13}. The predictions of the LD model show marked deviation from the experimental data, especially at the low A region. This explains the discrepancy observed in Fig.3  of Ref.~\cite{BOPO07} in which both the LD and the generalized mass formula were used.

\vspace{0.5cm}

\begin{table}[htbp]
\centering
\caption{Comparison of the  $\Lambda\Lambda$-separation energies (in MeV) of different $\Lambda\Lambda$-hypernuclei  obtained from the mass formula (SAM), experimental data ($B^{Exp}_{\Lambda\Lambda}$)~\cite{GAL11} and the shell model predictions ({$B^{SM}_{\Lambda\Lambda}$})~\cite{GAL11}.}
\tabcolsep11pt\begin{tabular}{lcccc}
\hline
\hline
\vspace{0.5cm}
Nucleus & { $B_{\Lambda\Lambda}$} & {$B^{Exp}_{\Lambda\Lambda}$}  & {$B^{SM}_{\Lambda\Lambda}$}\\
$^A_{\Lambda\Lambda}$Z & (SAM) &  \cite{GAL11}   & \cite{GAL11}\\
\hline
\hline
$^{	13}_{\Lambda\Lambda}$B 	& 22.91~	&~ $23.30\pm 0.70$~    & ~$23.21 \pm 0.21$	\\
$^{	12}_{\Lambda\Lambda}$B 	& 21.44~	&~ $20.60 \pm  0.74$~  & ~$20.85\pm 0.20$	\\
$^{	12}_{\Lambda\Lambda}$Be	& 22.21~      & ~$22.48\pm 1.21$~  & ~$20.72\pm 0.20$	\\
$^{	11}_{\Lambda\Lambda}$Be	& 20.16~      &~ $20.83\pm 1.27$~  & ~$ 18.40\pm 0.28$\\
$^{	10}_{\Lambda\Lambda}$Be	& 18.34~      & ~$14.94\pm 0.13$~  & ~$ 14.97\pm0.22$\\
$^{	 6}_{\Lambda\Lambda}$He	& ~~~7.12 ~ & ~~~$6.91\pm 0.16$~  &~~ $ 6.91 \pm 0.16$\\
\hline
\end{tabular} 
\label{table2}
\end{table}
\noindent 

\noindent

\begin{figure}[htbp]
  \centerline{\includegraphics[height=13.6cm,width=11.0cm]{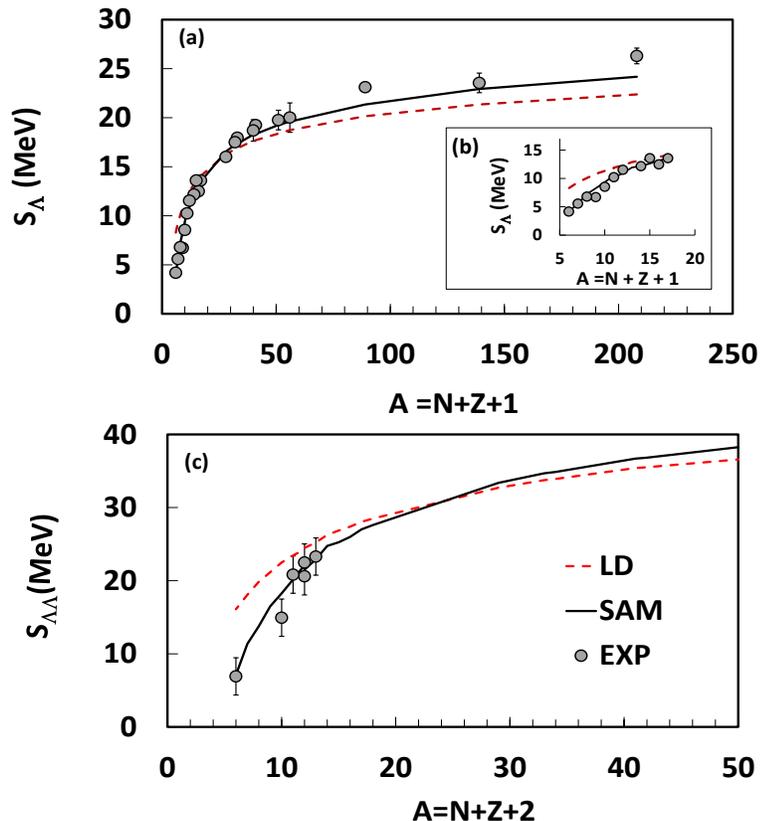}}
  \caption{Comparison of the experimental data of $\Lambda$- and $\Lambda\Lambda$-separation energies with the theoretical calculations using the liquid drop mass formula (LD)~\cite{BOPO07,BUY13} and the generalized mass formula (SAM)~\cite{SAM18} for (a) single-$\Lambda$ hypernuclei in wide mass region, (b) single-$\Lambda$ hypernuclei in low mass region (inset), and (c) double-$\Lambda$ hypernuclei.}
\end{figure}

\subsection {\bf{ The $\Lambda\Lambda$-Bond energy  ($\Delta B_{\Lambda\Lambda}$)}}
\label{sec:3}
\vspace{0.5cm}

\noindent
Inside a nucleus, a $\Lambda$-hyperon is found to show bonding with another $\Lambda$-hyperon and this bonding is found to be medium-dependent as it varies with the mass number A. The $\Lambda\Lambda$-bond energy is expected to shed some light on the nature of the in-medium strength of the~$\Lambda$-$\Lambda$ interaction.  

In Fig. 2 the results obtained with the mass formula (SAM) are  presented  with the experimental data and the predictions of QMF and RMF. From this figure the usefulness of the mass formula is evident as it can be used to evaluate the $\Lambda\Lambda$-bond energy in any nuclei over a wide mass range for any combination of neutrons and protons. The overall fit between the experimental and calculated $\Delta B_{\Lambda\Lambda}$ is good except near the low mass region.  The deviation at A=N+Z=4 arises from the under-prediction of the calculated $^4$He +$\Lambda$ binding energy. Measurement of $\Lambda$-separation energy for $^{11}_{\Lambda}$Be (Table 1) would provide another data point in this plot as $\Lambda\Lambda$-separation energy for $^{12}_{\Lambda\Lambda}$Be has already been measured (Table 2). Current data above A=4 have too large error bars. More accurate data with smaller error bars are needed to pinpoint the actual values of $\Delta B_{\Lambda\Lambda}$.

To further explore the medium effects for any specific hypernucleus, the $\Delta B_{\Lambda\Lambda}$ has been calculated for different nuclei with increasing neutron numbers. It is found that the bond-energy depends on the neutron number of the hypernucleus (Fig 3), and it decreases in value as the neutron number increases, which means that for the neutron-rich drip-line hypernuclei the $\Delta B_{\Lambda\Lambda}$ will be close to zero. This finding is interesting as it shows that the $\Lambda\Lambda$-bond energy  in a hypernucleus diminishes with both increasing proton and neutron numbers of the surrounding medium, indicating definite interplay between $\Lambda\Lambda$ and $\Lambda$-nucleon interactions.

\begin{figure}[htbp]
  \centerline{\includegraphics[height=10.0cm,width=9cm]{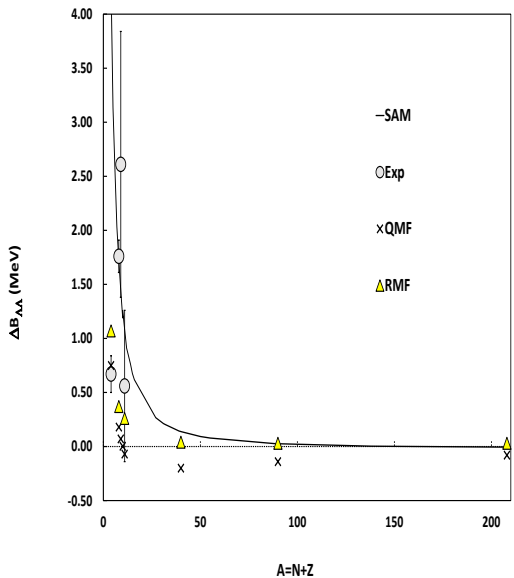}}
  \caption{Variation of $\Lambda$-$\Lambda$ bond energy ($\Delta B_{\Lambda\Lambda}$) with mass number A. Experimental values (Exp), and theoretical values calculated with the mass formula (SAM), QMF and RMF  are presented for comparison.}
\end{figure}
\vspace{0.5cm}

\begin{figure}[htbp]
  \centerline{\includegraphics[height=9.0cm,width=9cm]{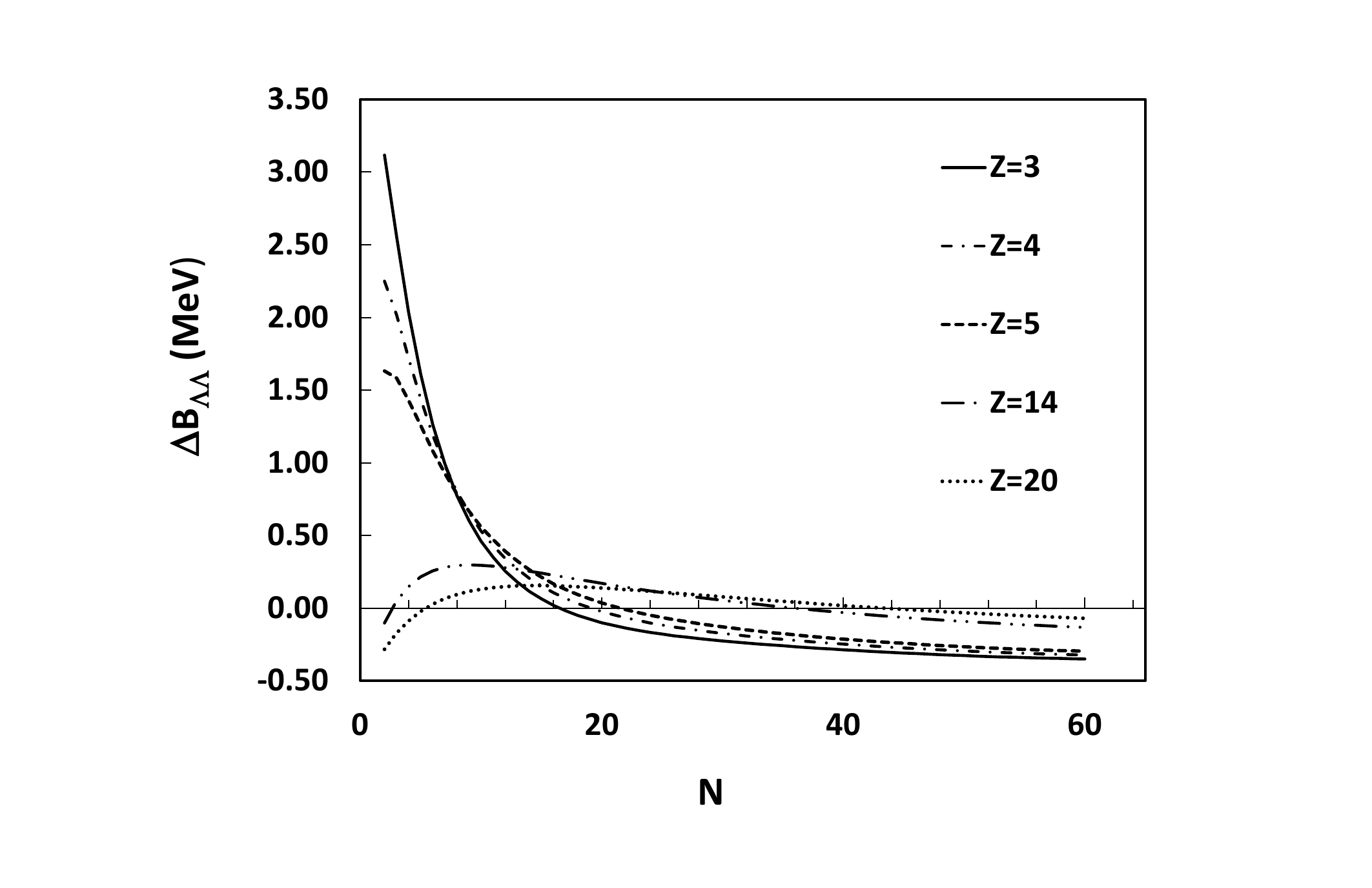}}
  \caption{Variation of $\Lambda$-$\Lambda$ bond energy ($\Delta B_{\Lambda\Lambda}$)  with the neutron number (N)}
\end{figure}

\vspace{0.8cm}

\subsection {\bf{ Charge symmetry breaking}}
\label{sec:4}
\vspace{0.5cm}
The $\Lambda$-hyperon has no isospin and no charge. The charge symmetry of strong interaction indicates that in mirror hypernuclei $\Lambda$p and $\Lambda$n interactions and their contributions in the $\Lambda$-binding energy should be identical. However, a comparison of the experimental $B_{\Lambda}$ values of the mirror-pair does not support this notion. The difference between those two values indicate the charge symmetry breaking (CSB) effect in $\Lambda$-hypernuclei. In the  mirror-pair, the mass number (A=N+Z+$n_\Lambda$) is the same for both the hypernuclei, while their charge numbers (Z) are interchanged with their neutron numbers (N). The respective cores of these hypernuclei are normal nuclei having the same charge as the corresponding hypernuclei, but  with  reduced mass number (A=N+Z) due to the absence of the $\Lambda$-hyperon. As the size of a nucleus depends on A, the Coulomb effect is larger in the core nuclei than the $\Lambda$-hypernuclei. On inspection of the Eqn.3 we find that in the  generalized mass formula,  the difference in $\Lambda$-hyperon binding energies, $\Delta B^c_{\Lambda}$(SAM), in mirror hypernuclei arises basically due to  the difference in the contributions of the Coulomb terms, not the CSB effect.  Nevertheless it is important as it gives an estimation of the Coulomb correction that needs to be used to extract the CSB effect from the experimental data.   When the negative Coulomb difference ($\Delta B^c_{\Lambda}$) exceeds the difference in lambda-nucleon interaction contributions ($\Delta$V$_{\Lambda N}$)  in binding energies, it results in negative values of ${\Delta}B_{\Lambda}^A$~\cite{GAL15}.

\begin{table}[htbp]
\centering
\caption{The binding energy difference $\Delta B_{\Lambda}$ (in MeV) in mirror nuclei. The $\Delta B^c_{\Lambda}$(SAM) values are the Coulomb energy difference from the generalized mass formula. Experimental data are taken from Table 1.}
\tabcolsep11pt\begin{tabular}{lcccc}
\hline
\hline
Mirror pair &  ${\Delta}B^c_{\Lambda}$ & ${\Delta}B_{\Lambda}^{Exp}$ & Ref.\\
   & (SAM) & &\\
\hline
\hline
$^{16}_\Lambda$O~-~$^{16}_\Lambda$N  &-0.086 &~ $-0.74\pm0.52$ & \cite{BOT17}	\\
$^{12}_\Lambda$C~-~$^{12}_\Lambda$B &~-0.091 ~ &~ $-0.61\pm0.20$	& \cite{BOT17} \\
$^{10}_\Lambda$B~-~$^{10}_\Lambda$Be  &-0.094  &~ $-0.41\pm0.24$ & \cite{BOT17}\\
$^{9}_\Lambda$B~-~$^{9}_\Lambda$Li  &-0.191 &~ $-0.24\pm0.23$ &  \cite{BOT17}\\
$^{8}_\Lambda$Be~-~$^{8}_\Lambda$Li &-0.097	&~~~~ $0.04\pm0.06$ & \cite{BOT17} \\
\hline
\end{tabular} 
\label{table3}
\end{table}
\noindent

\section{Summary}
	Newly available  experimental data on $\Lambda\Lambda$-hypernuclei are useful in providing information on the in-medium  $\Lambda$N, $\Lambda\Lambda$N and $\Lambda\Lambda$-interaction energies.   In this work we employ a generalized mass formula that gives excellent agreement with the experimental data of the binding energies of  single-$\Lambda$ and  double-$\Lambda$-hypernuclei. (It may be mentioned here that the same mass formula, without any change of parameters, reproduces the available Cascade ($\Xi^-$)-hyperon separation-energy data from different Cascade-hypernuclei). 

In the past, the $\Lambda\Lambda$-bond energies  were calculated for a few nuclei  in a quark mean field (QMF) model  and relativistic mean field (RMF) model~\cite{HU14}. The $\Lambda\Lambda$-bond energies calculated with the generalized mass formula~\cite{SAM18}  are in good agreement with the QMF and RMF predictions, as well as with the experimental data, except for the very light hypernuclei (Fig. 2). It is seen that the $\Lambda\Lambda$- interaction energy is high for low Z nuclei and diminishes rapidly as Z increases. In addition, we found a new feature not reported earlier, that even for a light hypernucleus with low Z-values, the $\Lambda\Lambda$-bond energy rapidly diminishes with the increasing neutron numbers (Fig. 3), and approaches zero for neutron-rich nuclei. This possibly reflects the relative strength and range of the in-medium $\Lambda$N,  $\Lambda\Lambda$ and $\Lambda\Lambda$N-interactions. To understand these intriguing aspects, more data on neutron-rich light hypernuclei are needed. 

The charge symmetry breaking effect is an interesting puzzle in hypernuclear physics. A considerable part of the difference in $\Lambda$-binding energies experimentally observed in mirror $\Lambda$-hypernuclei is known to arise from the hyperon-nucleon charge symmetry breaking effect. It is pertinent to note that the difference in $\Lambda$-binding energies in mirror $\Lambda$-hypernuclei evaluated using the generalized mass formula~\cite{SAM18} is not a measure of the charge symmetry breaking effect in hypernuclei. It is the Coulomb energy difference that is important as it can be utilized to determine the  Coulomb-corrected charge symmetry breaking effect in $\Lambda$-hypernuclei. 

The overall good agreement of this generalized mass formula with the experimental data shows that it can be used  in apriori estimation of the  hyperon(s) separation energy and  in-medium $\Lambda\Lambda$-bond energy  in the mass region not explored  through experiments so far. Thus it can provide a guidance for future experiments. Because of its simple formulation, it can be used as an input in multifragmentation production calculations~\cite{BOPO07,BUY13} as well as in fission calculations for hypernuclei.

\section{ACKNOWLEDGMENTS}
\vspace{0.5cm}
The authors acknowledge the support of the Department of Physics and Astronomy, Virginia Military Institute.

\end{document}